\documentstyle[aps,amsfonts,epsfig]{revtex}
\begin{document}
\draft
\twocolumn{
\title{Light propagation in non-trivial QED vacua}}
\author{Walter Dittrich and Holger Gies\thanks{E-mail address: 
holger.gies@uni-tuebingen.de}}

\address{Institut f\"ur theoretische Physik\\
          Universit\"at T\"ubingen\\
      Auf der Morgenstelle 14, 72076 T\"ubingen, Germany}
\date{}
\maketitle
\begin{abstract}
  Within the framework of effective action QED, we derive the light
  cone condition for homogeneous non-trivial QED vacua in the
  geometric optics approximation. Our result generalizes the ``unified
  formula'' suggested by Latorre, Pascual and Tarrach and allows for
  the calculation of velocity shifts and refractive indices for soft
  photons travelling through these vacua. Furthermore, we clarify the
  connection between the light velocity shift and the scale anomaly.
  This study motivates the introduction of a so-called effective
  action charge that characterizes the velocity modifying properties
  of the vacuum. Several applications are given concerning vacuum
  modifications caused by, e.g., strong fields, Casimir systems and
  high temperature.
\end{abstract}
\nopagebreak
\section{Introduction}
The vacuum considered as a medium has become a popular picture in
quantum field theory. With re\-ser\-vations due to the lack of
understanding of non-perturbative vacuum phenomena, it is astonishing
that analogies between the quantum vacuum and classical media are
frequently useful.

One particular example is represented by the propagation of light in a
vacuum which is modified by various external environments, e.g.,
electromagnetic (EM) fields, temperature, geometric boundary
configurations, gravitational background and non-trivial topologies.
The concept of drawing the analogy is common to all of these cases:
vacuum polarization allows the photon to exist as a virtual
$e^+e^-$-pair on which the various vacuum modifications can act. Under
certain assumptions, this influence on the loop process can
effectively be described by an immediate influence of a (generally
non-linear) medium on the photon itself, e.g., by refractive indices.
This program was carried out among others by Adler \cite{adler1},
Brezin and Itzykson \cite{brezin} for magnetic fields, by Drummond and
Hathrell \cite{drummond} for gravitation, and by Scharnhorst
\cite{scharnhorst1} and Barton \cite{barton1} for a Casimir
configuration. Further important examples are found in refs.
\cite{daniels1,daniels2,latorre,tsai1}.

A new physical insight into the phenomenon of photon propagation in
non-trivial vacua has been given by Latorre, Pascual an Tarrach
\cite{latorre}. Comparing the known velocity shifts arising from
different vacuum modifications, they were able to identify an
intriguing general, so-called ``unified'' formula covering all these
cases\footnote{In fact, the results of \cite{tsai1} cannot be embedded
  in the ``unified formula''. The solution to this problem is an aim
  of the present work.}. They concluded that the polarization and
direction averaged velocity shift is related to the (renormalized)
background energy density $u$ with a ``universal'' numerical
coefficient

\begin{equation}
\delta \bar{v}=-\frac{44}{135} \frac{\alpha^2}{m^4}\, u\, \label{1}
\end{equation}
where $m$ denotes the electron mass and $\alpha\simeq 1/137$. (In the
case of gravitation, one $\alpha$ has to be replaced by the
combination $(G_{\text{N}}m^2)$ involving Newton's constant). However,
a complete derivation of the ``unified formula'' has not been given up
to now.

In the case of gravitation, light was shed on the problem by Shore
\cite{shore} who proved a polarization sum rule that represents a
generalization of eq.(\ref{1}). Furthermore, he pointed out that the
``universal'' coefficient in eq.(\ref{1}) can be related to the trace
anomaly of the energy-momentum tensor in the case of weak EM
background fields.

One of the most remarkable features concerning vacuum induced velocity
shifts certainly is the fact that $\delta \bar{v}>0$ is not
intrinsically forbidden in quantum field theories. This seems to offer
the possibility of superluminal propagation, e.g., in curved spaces
and Casimir vacua. Both examples share the property of a possible
negative energy density $u$ in eq.(\ref{1}).

The two questions, whether the signal ($=$wave front) velocity indeed
exceeds $c$ and whether superluminal propagation is observable in
principle, could be resolved by calculating the velocity shift in the
infinite frequency limit. But this is presently out of reach, because
a resummation of the derivative expansion has to be achieved. However,
without being able to answer these questions, let us just say that we
find no grounds for violation of (micro-) causality in accordance with
\cite{drummond,latorre,shore,scharnhorst2}. For a causality violation,
a space-like signal {\em and} Lorentz invariance (in the gravitational
case: strong principle of equivalence) are necessary conditions. The
latter is explicitly violated in the above-mentioned examples. For an
excellent discussion, the reader is referred to the work of Shore
\cite{shore}.

In the present work, we confine ourselves to the case of non-trivial
vacua modified by QED phenomena. Within the effective action approach
\cite{dittreut}, we derive a covariant light cone condition in sect.
II which turns out to be a generalization of the ``unified formula''.
The necessary assumptions are analysed in detail.

In this framework, we are able to clarify the relation between the
velocity shift and the trace anomaly in sect. III. Our findings do not
unveil a natural and physically meaningful connection. An alternative
physical picture of the ``universal'' pre-factor is given instead
which is called: effective action charge. Several applications of our
light cone condition concerning EM fields, Casimir configurations and
temperature are elaborated on in sect. IV. In the low-energy domain,
we can easily recover all of the well-known results described by the
``unified formula''. However, the ``universal constant'' turns out to
be neither constant nor universal when we drop the low-energy
restriction. Instead, the concept of an effective action charge
provides for an intuitive understanding of the velocity shifts at
arbitrary energies.

Conclusions are drawn in sect. V.

\section{The Light Cone Condition}
Consider light propagation in a non-trivial QED vacuum (we will
specify this terminology soon) characterized by a certain energy
scale.  Suppose that there exists an effective action which takes into
account any QED quantum phenomena on higher scales and hence provides
for an exact description of the propagation. In principle, this
effective action will depend on any gauge and Lorentz invariant scalar
which we can construct. Throughout the paper, we will stick to the
following essential assumptions:

\noindent
1) The propagating photons characterized by $f^{\mu\nu}$ are
considered to be soft. This 
is equivalent to calculating the properties of the
vacuum in the limit $\omega/m \ll 1$ where the scale is set by the
Compton wavelength.

\noindent
2) The vacuum modification is homogeneous in space and time (but not
necessarily isotropic).

Referring to these assumptions, we can neglect any term in the
effective action that involves derivatives of the field, since a
derivative either acting on the background field vanishes (assumption
2)) or acting on the photon field $f^{\mu\nu}$ contributes terms of
the order ${\cal O} (\omega^2/m^2)$ to the equation of motion. In the
latter case, it is negligible because of assumption 1).

We furthermore assume that

\noindent
3) Vacuum modifications caused by the propagating light itself are
negligible.

Assumption 3) justifies a linearization of the equations of motion
with respect to $f^{\mu\nu}$ but does not stand on the same footing as
the former assumptions, since it is not essential for the formalism.
Note that we do {\em not} demand that the deviation from the Maxwell
Lagrangian should be small, corresponding to small vacuum
modifications.\footnote{In principle, the Lagrangian ${\cal L}$ can
  contain imaginary parts indicating the instability of the modified
  vacuum. In the following, it is understood that we take into account
  only the real part of ${\cal L}$ which is solely responsible for the
  field equations.}

Since it is unwieldy to establish a general formalism for arbitrary
numbers of Lorentz vectors and tensors characterizing the vacuum, we
first consider a vacuum only modified by EM fields. Hence, the
dynamical building blocks of the effective action which respect
Lorentz and gauge invariance are given by the field strength tensor
and its dual

\begin{mathletters}
\label{2}
\begin{eqnarray}
F^{\mu\nu}&=&\partial^\mu A^\nu-\partial^\nu A^\mu\, \label{2a}\\
^\star\! F^{\mu\nu}&=& \case{1}{2} \epsilon^{\mu\nu\alpha\beta}
  F_{\alpha\beta}\, .\label{2b}
\end{eqnarray}
\end{mathletters}
The lowest-order linearly independent scalars are

\begin{mathletters}
\label{3}
\begin{eqnarray}
x&:=&\case{1}{4}F_{\mu\nu}F^{\mu\nu} =
  \case{1}{2}(\bbox{B}^2-\bbox{E}^2) \, \label{3a}\\
y&:=&\case{1}{4}F_{\mu\nu}\, ^\star\! F^{\mu\nu} = 
  \bbox{E\cdot B}\, , \label{3b}
\end{eqnarray}
\end{mathletters}
The normalization is chosen in such a way that the Maxwell Lagrangian
can be written ${\cal L}_{\text{M}}=-x$.\footnote{$x$ and $y$ are
  usually called ${\cal F}$ and ${\cal G}$. We do not follow this
  convention for reasons of simplicity.}  By taking advantage of the
antisymmetry of $F^{\mu\nu}$ and by virtue of the relations
\cite{schwinger}

\begin{mathletters}
\label{4}
\begin{eqnarray}
F^{\mu\alpha} F^\nu_{\,\,\,\alpha} -\, ^\star\! F^{\mu\alpha}\,
  ^\star\! F^{\nu}_{\,\,\,\alpha} &=& 2\, x\, g^{\mu\nu}\,
  ,\label{4a}\\ 
F^{\mu\alpha}\, ^\star\! F^{\nu}_{\,\,\,\alpha} =\, ^\star\!
  F^{\mu\alpha} F^\nu_{\,\,\,\alpha} &=& y\, g^{\mu\nu}\, ,\label{4b}
\end{eqnarray}
\end{mathletters}
using the metric $g=\text{diag} (-,+,+,+)$, it is easy to verify (i)
the vanishing of odd-order invariants and (ii) that invariants of
arbitrary order can be reduced to expressions only involving $x^n y^m$
where $n,m=0,1,2\dots$. Besides, note that parity invariance demands
for $m$ to be even.

%
%
Consequently, the complete effective action becomes extremely
simplified, turning out to be a function of $x$ and $y$ only. The
corresponding Lagrangian reads

\begin{equation}
{\cal L}={\cal L}(x,y)\, . \label{6}
\end{equation}
We obtain the equations of motion from ${\cal L}$ by variation

\begin{eqnarray}
0&=&\partial_\mu \frac{\partial {\cal L}}{\partial (\partial_\mu
  A_\nu)} -\frac{\partial {\cal L}}{\partial A_\mu} \nonumber\\
&=&\partial_\mu \bigl( \partial_x {\cal L}\, F^{\mu\nu} +\partial_y
  {\cal L}\, ^\star\! F^{\mu\nu} \bigr)\, \label{7}
\end{eqnarray}
where $\partial_x,\partial_y$ denote the partial derivatives with
respect to the field strength invariants (\ref{3}) (and should not be
confused with space-time derivatives $\partial_\mu$).

If we take advantage of the Bianchi identity while moving
$\partial_\mu$ to the right, we arrive at

\begin{equation}
0=(\partial_x {\cal L})\, \partial_\mu F^{\mu\nu} +\left(\case{1}{2}
M^{\mu\nu}_{\alpha\beta}\right)\, \partial_\mu F^{\alpha\beta}\, ,
\label{8}
\end{equation}
where $M^{\mu\nu}_{\alpha\beta}$ is given by

\begin{eqnarray}
M^{\mu\nu}_{\alpha\beta}&:=& F^{\mu\nu}F_{\alpha\beta}\, (\partial^2_x
  {\cal L}) +\, ^\star\! F^{\mu\nu}\, ^\star\! F_{\alpha\beta}\,
  (\partial^2_y {\cal L}) \nonumber\\
&& +\partial_{xy}{\cal L}\, \bigl( F^{\mu\nu}\, ^\star\! 
  F_{\alpha\beta}+\, ^\star\! F^{\mu\nu} F_{\alpha\beta} \bigr)\,
  .\label{9} 
\end{eqnarray}
Note that $M$ is antisymmetric in the upper as well as the lower
indices: $M^{\mu\nu}_{\alpha\beta}=-M^{\nu\mu}_{\alpha\beta}=
M^{\nu\mu}_{\beta\alpha}$.

In general, $F^{\mu\nu}$ contains background fields
$F_{\text{B}}^{\mu\nu}$ and the propagating photon field $f^{\mu\nu}$.
According to assumption 2), the derivative acting on
$F_{\text{B}}^{\mu\nu}$ vanishes

\begin{equation}
\partial_\mu F^{\lambda\kappa}=\partial_\mu f^{\lambda\kappa}\,
.\label{10}
\end{equation}
Inserting eq.(\ref{10}), eq.(\ref{8}) yields in Fourier space

\begin{equation}
0=(\partial_x {\cal L})\, k_\mu f^{\mu\nu} +\left(\case{1}{2}
M^{\mu\nu}_{\alpha\beta}\right)\, k_\mu f^{\alpha\beta}\, .\label{11}
\end{equation}
Introducing a gauge potential $a^\mu$ for the propagating field
$f^{\mu\nu}$, we may write

\begin{equation}
f^{\mu\nu}=k^\mu a^\nu- k^\nu a^\mu= a\, (k^\mu \epsilon^\nu -k^\nu
\epsilon^\mu )\, ,\label{12}
\end{equation}
where $a:=\sqrt{a^\mu a_\mu}$ and $\epsilon^\mu= a^\mu/a$. Here, the
polarization vectors $\epsilon^\mu$ are normalized to 1.

Establishing the Lorentz gauge $k_\mu \epsilon^\mu=0$, we get

\begin{equation}
0=(\partial_x {\cal L})\, k^2 \epsilon^\nu +
M^{\mu\nu}_{\alpha\beta}\, k_\mu k^\alpha \epsilon^\beta\,
,\label{13}
\end{equation}
where we used the antisymmetry of $M^{\mu\nu}_{\alpha\beta}$.

The next important step is to multiply eq.(\ref{13}) by $\epsilon_\nu$
and average over polarization states according to the well-known rule

%
%
\begin{equation}
\sum_{\text{pol.}} \epsilon^\beta \epsilon^\nu \to g^{\beta\nu}
\, ,\label{14}
\end{equation}
where the additional terms on the right-hand side of (\ref{14}) vanish
with the aid of the antisymmetry of $M^{\mu\nu}_{\alpha\beta}$.  We
find for eq.(\ref{13})

\begin{equation}
0=2(\partial_x {\cal L})\, k^2 +M^{\mu\nu}_{\alpha\nu}\, k_\mu k^\alpha 
\, .\label{15}
\end{equation}
%
%
Equation (\ref{15}) already represents a light cone condition and
actually indicates that the familiar $k^2=0$ will in general not hold
for arbitrary Lagrangians. Our final task is to put
$M^{\mu\nu}_{\alpha\nu}$ in a convenient shape. Using the powerful
relations (\ref{4}), we obtain

\begin{equation}
M^{\mu\nu}_{\alpha\nu}=2\left[ \case{1}{2} F^{\mu\nu}F_{\alpha\nu}
(\partial^2_x +\partial^2_y){\cal L} +\delta^\mu_\alpha (y
\partial_{xy}{\cal L} -x\partial^2_y {\cal L}) \right]
\, .\label{16}
\end{equation}
Introducing the Maxwell energy-momentum tensor

\begin{equation}
T^\mu_{\,\,\,\, \alpha}=F^{\mu\nu} F_{\alpha\nu} -x
\,\delta^\mu_\alpha\, ,\label{17}
\end{equation}
this leads to

\begin{eqnarray}
M^{\mu\nu}_{\alpha\nu}=2\bigl[&&\case{1}{2}T^\mu_{\,\,\,\, \alpha}
  (\partial^2_x +\partial^2_y){\cal L} \nonumber\\
&&+\delta^\mu_\alpha \left(
  \case{1}{2} x (\partial^2_x -\partial^2_y){\cal L} +y
  \partial_{xy}{\cal L}\right)\bigr]\, .\label{18}
\end{eqnarray}
However, the Maxwell energy-momentum tensor in general is devoid of
any physical meaning, since we are simply not dealing with the Maxwell
Lagrangian. The right quantity to deal with is therefore the vacuum
expectation value (VEV) of the energy-momentum tensor defined
by\footnote{Note that the variation with respect to the metric tensor
  is just a trick to calculate the symmetric energy-momentum tensor.
  With some care, the same result can be obtained by canonical
  methods.}

\begin{equation}
\langle T^{\mu\nu}\rangle:= \frac{2}{\sqrt{-g}} \frac{\delta
  \Gamma}{\delta g_{\mu\nu}} \quad, \, \Gamma:= \int d^4x\, \sqrt{-g}
  \, {\cal L}\, , \label{19}
\end{equation}
where $\Gamma$ denotes the effective action. Performing the
calculation, we arrive at

\begin{equation}
\langle T^{\mu\nu}\rangle_{xy}=-T^{\mu\nu} (\partial_x {\cal L})
+ g^{\mu\nu}\, ({\cal L} -x\partial_x {\cal L} -y\partial_y {\cal
  L})\, .\label{20}
\end{equation}
%
%
Solving eq.(\ref{20}) for $T^{\mu\nu}$ and inserting into
eq.(\ref{18}), we can present $M^{\mu\nu}_{\alpha\nu}$ in its final
form

%
%
\begin{eqnarray}
M^{\mu\nu}_{\alpha\nu}=2\biggl[&&-\frac{1}{2}\frac{(\partial^2_x\!
  +\!\partial^2_y){\cal L}}{\partial_x {\cal L}}  \langle
  T^{\mu}_{\,\,\,\,\alpha}\rangle_{xy}+\delta^\mu_\alpha
  \Bigl(\case{1}{2} x(\partial^2_x\!-\!\partial^2_y){\cal L}
  \nonumber\\ 
&&\, +y \partial_{xy}{\cal L} +\frac{\frac{1}{2}(\partial^2_x\!
  +\!\partial^2_y){\cal L}}{\partial_x{\cal L}}({\cal L}\!-\!x
  \partial_x {\cal L}\! -\!y\partial_y {\cal L})\Bigr)\biggr]\,
  .\nonumber\\ 
&&\label{21}
\end{eqnarray}
Substituting $M^{\mu\nu}_{\alpha\nu}$ into eq.(\ref{15}), we end up
with the desired light cone condition for EM field modified vacua
fulfilling the above-mentioned assumptions

\begin{equation}
k^2\, =\, Q\, \langle T^{\mu\nu}\rangle_{xy}\, k_\mu k_\nu\, ,\label{23d}
\end{equation}
where

%
%
\begin{equation}
Q=\frac{\frac{1}{2} (\partial^2_x +\partial^2_y){\cal L}}
{{\scriptstyle \Bigl[ \!(\partial_x\!{\cal L})^2\!+(\partial_x\!{\cal
    L})\!\bigl(\!\case{x}{2} (\partial^2_x\! -\partial^2_y)+y
  \partial_{xy}\!\bigr)\!{\cal L}\!+\frac{1}{2}\! (\partial^2_x
  \!+\partial^2_y){\cal L}(1\! -x\partial_x \!-y\partial_y)\!
  {\cal  L} \Bigr]}}. \label{23}
\end{equation}
To extend the validity of the light cone condition to arbitrary
non-trivial vacua, we have to take the vacuum expectation value of
eq.(\ref{23d}) with respect to the additional vacuum modifications
parametrized by the (collective) label $z$

\begin{equation}
k^2\, =\, _z\langle 0|\,Q\, \langle T^{\mu\nu}\rangle_{xy}\,
|0\rangle_z \, k_\mu k_\nu\, .\label{23a}
\end{equation}
Inserting a complete set of intermediate states, we obtain

\begin{equation}
k^2\, =\sum_i\,\!  _z\langle 0|\,Q\,|i\rangle_{z\quad\!\!\!\!\!
  z}\!\langle i| \langle T^{\mu\nu}\rangle_{xy}\, |0\rangle_z \, k_\mu
  k_\nu\, .\label{23b} 
\end{equation}
In the following, we consider the vacuum to behave as a {\em passive}
medium in which the EM fields and the further vacuum modifications $z$
remain in a state of static equilibrium. Since $Q$ solely depends on
$x$ and $y$ (via ${\cal L}(x,y)$), this {\em assumption of passivity}
leads to 

\begin{equation}
  _z\langle 0|\,Q\,|i\rangle_{z}=\langle Q\rangle_z\, \delta_{0i}\,
  .\label{23c} 
\end{equation}
Equation (\ref{23c}) states that the vacuum exhibits no back-reaction
caused by the EM fields while switching on $z$.

$Q$ depends functionally on ${\cal L}(x,y)$, which is, as usual,
defined via the functional integral over the fluctuating fields.
Taking the expectation value of $Q$ hence leads back to integrating
over the field configurations which respect the modified vacuum. E.g.,
if the modification $z$ imposes boundary conditions on the fields, the
functional integral has to be taken over the fields which fulfil these
boundary conditions. Therefore, taking the VEV of $Q$ defines the new
effective Lagrangian characterizing the complete non-trivial vacuum

\begin{equation}
\langle Q\rangle_z=\langle Q({\cal L}(x,y))\rangle_z=Q({\cal
  L}(x,y;z))\, .\label{23e}
\end{equation}
We finally arrive at the light cone condition for arbitrary
homogeneous non-trivial vacua

\begin{equation}
k^2=Q(x,y,z)\, \langle T^{\mu\nu} \rangle_{xyz}\, k_\mu k_\nu\,
.\label{22}
\end{equation}
Remember that the validity of the light cone condition eq.(\ref{22})
is not restricted to results of perturbation theory or only small
modifications of the Maxwell Lagrangian. It is an exact statement in
the sense of effective theories.

Now, the terminology ``modified QED vacuum'' should be clarified: from
the derivation of the light cone condition, it is obvious that the
implicit space-time dependence of ${\cal L}$ should only be contained
in the field variables. Furthermore, the vacuum has to fulfil the
demand for passivity. Otherwise the light cone condition (\ref{22})
only represents a zeroth order approximation of the infinite sum over
intermediate states in eq.(\ref{23c}). 

As a third remark, we want to point out that the sum over polarization
states is not necessary for the derivation of a light cone condition.
By summing, we even exclude the study of birefringence from the
formalism which is certainly the most important experimental
application \cite{bakalov1,bakalov2,iacopini}. But for a projection on
the polarization eigenstates, the $y^n$-terms have to be rewritten in
terms of the field strength tensor which is practically impossible for
arbitrary ${\cal L}$.

In the remainder of the section, we calculate further representations
of eq.(\ref{22}) by choosing a certain reference frame and introducing

\begin{equation}
\bar{k}^\mu=\frac{k^\mu}{|\bbox{k}|} =\left(\frac{k^0}{|\bbox{k}|}
  ,\bbox{\hat{k}} \right) =:(v, \bbox{\hat{k}})\, ,\label{24}
\end{equation}
where we defined the phase velocity by $v:=k^0/|\bbox{k}|$. For
eq.(\ref{22}), we obtain

\begin{equation}
v^2=1-Q\, \langle T^{\mu\nu}\rangle \bar{k}_\mu\bar{k}_\nu\,
.\label{25}
\end{equation}
Equation (\ref{25}) clearly demonstrates that the light cone condition
is a generalization of the ``unified formula'' of Latorre, Pascual and
Tarrach \cite{latorre}.

In general, the $Q$-factor will depend on all the variables and
parameters of ${\cal L}$ and hence will naturally be neither universal
nor constant. Besides, the daunting structure of the $Q$-factor will
simplify in the case of small corrections to ${\cal L}_{\text{M}}$. As
will be shown in sect. IV, the denominator then reduces to 1.

Another representation of the light cone condition is found by
averaging over propagation directions, i.e., integrating over
$\bbox{\hat{k}} \in S^2$

\begin{equation}
v^2=\frac{1-Q\bigl(\case{1}{3} \langle T^{00}\rangle
  +\case{1}{3}\langle T^{\alpha}_{\,\,\,\alpha}\rangle\bigr)}{1+Q\, \langle
  T^{00}\rangle}\, .\label{26}
\end{equation}
For $Q \langle T^{00}\rangle\ll 1$ and $\langle
T^{\alpha}_{\,\,\,\alpha}\rangle$ being even of lower order, this
reduces to

\begin{equation}
v^2=1-\frac{4}{3}\, Q\, \langle T^{00}\rangle= 1-\frac{4}{3}\, Q\, u\,
,\label{27}
\end{equation}
where $u$ denotes the (renormalized) energy density of the modified
vacuum.

\section{velocity shift and scale anomaly}
In his paper, Shore \cite{shore} suggested a deeper
connection between the velocity shift and the scale anomaly. For the
Heisenberg-Euler Lagrangian, he showed that the coefficients of the
$x^2$ and $y^2$ terms in the scale anomaly are precisely those
appearing in the velocity shift for the different polarization
states. 

Within the framework developed so far, we will attempt to clarify the
relation between the scale anomaly and the velocity shift. Therefore,
we have to investigate whether the terms in the $Q$-factor can be
expressed in terms of the anomaly. For reasons of simplicity, we limit
this consideration to the case of a purely EM field modified vacuum.
From eq.(\ref{20}), we can read off the scale anomaly

\begin{equation}
\langle T^{\alpha}_{\,\,\,\alpha}\rangle=4 ({\cal L} -x \partial_x
{\cal L} -y \partial_y {\cal L} )\, . \label{28}
\end{equation}
By differentiation, we find

\begin{mathletters}
\label{29}
\begin{eqnarray}
\partial_x \langle T^{\alpha}_{\,\,\,\alpha}\rangle&=& -4
  (x\partial^2_x {\cal L} +y\partial_{xy}{\cal L} )\, ,\label{29a}\\
\partial_y \langle T^{\alpha}_{\,\,\,\alpha}\rangle&=& -4
  (y\partial^2_y {\cal L} +x\partial_{xy}{\cal L} )\, .\label{29b}
\end{eqnarray}
\end{mathletters}
From eqs.(\ref{29}) immediately follows

\begin{equation}
(\partial^2_x+\partial^2_y){\cal
  L}=-\bigl(\case{y}{x}+\case{x}{y}\bigr) \partial_{xy}{\cal L}
  -\case{1}{4} \bigl(\case{1}{x}\partial_x +\case{1}{y}\partial_y
  \bigr) \langle T^{\alpha}_{\,\,\,\alpha}\rangle\, .\label{30}
\end{equation}
This expression is proportional to the numerator of the $Q$-factor
eq.(\ref{23}). Using similar techniques, we can also rewrite the
denominator, but the result is not very illuminating

%
%
\begin{eqnarray}
\text{denom.}(Q)&=& (\partial_x {\cal L})^2 -(\partial_x {\cal L}) \bigl(
  \case{x}{2} (\partial^2_x+\partial^2_y){\cal L}+\case{1}{4}
  \partial_x \langle T^{\alpha}_{\,\,\,\alpha}\rangle\bigr)\nonumber\\
&& +\case{1}{8}\bigl[(\partial^2_x+\partial^2_y){\cal L}\bigr]
  \langle T^{\alpha}_{\,\,\,\alpha}\rangle\, .\label{30b}
\end{eqnarray}
Fortunately, approximating eq.(\ref{30b}) by 1 will be appropriate to
the applications of sect. IV.

It is already obvious from eq.(\ref{30}) that there is no immediate
connection between $\langle T^{\alpha}_{\,\,\,\alpha}\rangle$ and the
velocity shift eq.(\ref{25}). The findings of Shore arise from the
special structure of the Heisenberg-Euler Lagrangian where
$\partial_{xy} {\cal L}=0$. In general, higher-order mixed terms are
not forbidden by gauge, Lorentz or parity invariance. Referring to
eq.(\ref{30}), the introduction of the scale anomaly appears to be
artificial rather than interpretable. Even Shore's conjecture that the
sign of the scale anomaly is linked to the sign of the velocity shift
cannot be maintained.

Instead, we favour the pure effective action formulation, i.e., the
left-hand side of eq.(\ref{30}), since it offers a new intuitive
picture. Referring to eq.(\ref{25}), the value and sign of the
velocity shift result from the competition between the VEV of the
energy-momentum tensor and the $Q$-factor. Both are a priori neither
positive nor bounded by symmetry principles. Let us restrict the
following investigation to the case of small corrections to the
Maxwell Lagrangian, i.e., 

\begin{equation}
Q\simeq \frac{1}{2}(\partial^2_x+\partial^2_y){\cal L} \quad
\Longrightarrow \bbox{\nabla}^2 {\cal L}=2\,Q \, .\label{31}
\end{equation}
Due to the similarity to the (2-D) Poisson equation, we will call $Q$
from now on the {\em effective action charge} in field space. The
classical vacuum ${\cal L}_{\text{M}}=-x$ is {\em uncharged} and hence
$v=1$. As we will soon demonstrate, the pure QED vacuum has a small
positive charge at the origin in field space ($x=y=0$). For increasing
field strength, $\langle T^{\mu\nu}\rangle$ certainly also increases
without upper bound, so we expect $Q$ to decrease in order to produce
no unphysical velocity shift $>1$. It is therefore reasonable to
presume localized effective action charge distributions centred upon
the origin in field space. The results of sect. IV will confirm this
charge-like picture. 

\section{applications of the light cone condition}
Up to now, the light cone condition might be regarded as a nice frame
without a picture enclosed, since it is much easier to talk about
all-loop or non-perturbative effective actions than to calculate one. 

Indeed, the effective actions which we are going to insert will not
reach beyond two-loop order. Their general structure can be
characterized by

\begin{equation}
{\cal L}={\cal L}_{\text{M}} +{\cal L}_{\text{c}} \qquad; \quad
\frac{{\cal L}_{\text{c}}}{{\cal L}_{\text{M}}} \ll 1\, ,\label{32}
\end{equation}
where ${\cal L}_{\text{c}}$ contains the correction terms. 

Regarding the denominator expression of the effective action charge
(\ref{30b}), the scale anomaly $\langle
T^{\alpha}_{\,\,\,\alpha}\rangle$ is of the same order as ${\cal
  L}_{\text{c}}$. Hence, eq.(\ref{30b}) simply reduces to

\begin{equation}
\text{eq.(\ref{30b})}=1+{\cal O}({\cal L}_{\text{c}})\, ,\label{33}
\end{equation}
and the approximation $Q=\frac{1}{2}\bbox{\nabla}^2 {\cal L}=
\frac{1}{2}\bbox{\nabla}^2 {\cal L}_{\text{c}}$ is justified.

\subsection{Weak EM Fields}
According to the authors of ref. \cite{ritus}, the two-loop corrected
Hei\-sen\-berg-\-Euler Lagrangian (weak-field limit of the complete
one-loop approximated effective QED Lagrangian) reads

\begin{equation}
{\cal L}= -x+c_1\, x^2+ c_2\, y^2\, ,\label{34}
\end{equation}
where

\begin{mathletters}
\label{35}
\begin{eqnarray}
c_1&=&\frac{8\alpha^2}{45 m^4} \left( 1+\frac{40}{9}\frac{\alpha}{\pi}
  \right) \, , \label{35a}\\
c_2&=&\frac{14\alpha^2}{45 m^4} \left(
  1+\frac{1315}{252}\frac{\alpha}{\pi} \right) \, . \label{35b}
\end{eqnarray}
\end{mathletters}
With the aid of the light cone condition eq.(\ref{27}), we immediately
obtain for the polarization and propagation direction averaged velocity
($v\equiv \sqrt{\bar{v^2}}$)

\begin{eqnarray}
Q&=&c_1+c_2 \, ,\label{36vor}\\
\leadsto v&=&1-\frac{4\alpha^2}{135 m^4} \left(11+\frac{1955}{36}
  \frac{\alpha}{\pi} \right) \!\left[ \frac{1}{2} \bigl(
  \bbox{E}^2+\bbox{B}^2 \bigr) \right]\, .\label{36}
\end{eqnarray}
In the well-known one-loop part of eq.(\ref{36}), we can identify the
factor of $\frac{44\alpha^2}{135m^4}$ as the ``universal constant'' of
the ``unified formula'' eq.(\ref{1}). At this stage, it is already
understandable that all of the known QED induced velocity shifts share
this universal factor in the low-energy limit, since they are all
based on the Heisenberg-Euler Lagrangian. Even the results in
gravitation involve the same $e^+ e^-$-loop calculation (of course, in
a curved space-time \cite{drummond}).\footnote{In particular, there is
  nothing mysterious about the factor  11 as it is sometimes found
  in the literature.}
It is furthermore obvious that the two-loop correction
$\frac{1955}{36}\frac{\alpha}{\pi}$ is as universal as the
number 11. (Note that modifications from the denominator of $Q$
eq.(\ref{30b}) contribute to the order ${\cal O}(\alpha^4))$.

\subsection{Strong Magnetic Fields}
Since the Heisenberg-Euler Lagrangian eq.(\ref{34}) represents a weak
field limit, eq.(\ref{36vor}) denotes the value of the effective
action charge at the origin in field space ($x,y=0$). In this
subsection, we analyse the form of $Q$ along the positive $x$-axis
(pure magnetic fields). As our starting point, we use Schwinger's
famous formula for the one-loop effective QED Lagrangian
\cite{schwinger} 

%
%
\begin{eqnarray}
{\cal L}_{\text{c}}\!=-\frac{1}{8\pi^2}\!\!
  \int\limits_0^{\text{i}\infty}
  \!\!\frac{ds}{s^3}&&\text{e}^{\!-m^2\!s}\!\biggl[(es)^2|y|
  \coth\!\Bigl(\!es\bigl(\!
  \sqrt{x^2\!+\!y^2}\!+x\bigr)\!^{\case{1}{2}}\!\Bigr) \nonumber\\
&&\times\cot\!\Bigl(\! es\bigl(\!\sqrt{x^2\!+\!y^2}
  \!-x\bigr)\!^{\case{1}{2}}\!\Bigr)\!-\frac{2}{3}(es)^2
  x-1\!\biggr]\!. \nonumber\\ 
&&\label{37}
\end{eqnarray}
It is understood that the convergence is implicitly ensured by the
prescription $m^2 \to m^2-\text{i}\epsilon$. (Note that we have not
performed a proper time Wick rotation yet.) 

It will be useful to reparametrize the field space with new
coordinates

\begin{mathletters}
\label{38}
\begin{eqnarray}
a:=\bigl(\sqrt{ x^2+y^2} +x\bigr)^{\case{1}{2}}\, &,&\,
  b:=\bigl(\sqrt{   x^2+y^2} -x\bigr)^{\case{1}{2}}\, ,\label{38a} \\
\Longrightarrow \qquad |y|=ab\quad &,&\, x=\case{1}{2}(a^2-b^2)\,
  .\label{38b}
\end{eqnarray}
\end{mathletters}
The Laplacian in terms of $a$ and $b$ reads

\begin{equation}
\bbox{\nabla}^2 =\frac{1}{a^2+b^2} \bigl( \partial_a^2+ \partial_b^2
\bigr)\, .\label{39}
\end{equation}
For the term in the square brackets in eq.(\ref{37}), we easily find

%
%
\begin{eqnarray}
\bbox{\nabla}^2 \Bigl[ \cdots \Bigr]&=&\frac{(es)^2}{a^2+b^2} \bigl(
  \partial_a^2+ \partial_b^2 \bigr)\bigl[ ab\coth esa\,\cot esb\bigr]
  \nonumber\\
&=&\frac{2(es)^2}{a^2+b^2}\!\left[ \frac{esb\cot esb}{\sinh^2 esa}
  (esa \coth esa -1) \right.\label{40}\\
&&\left. \qquad\quad\,\, +\frac{esa \coth esa}{\sinh^2 esb} (esb \cot
  esb -1) \right] .\nonumber
\end{eqnarray}
Confining ourselves to purely magnetic fields ($x=\case{1}{2} B^2$,
$y=0$ $\Rightarrow$ $b=0$, $a=|\bbox{B}|$), we obtain

%
%
\begin{equation}
\text{eq.(\ref{40})}\to \frac{2(es)^2}{a^2} \left[ \frac{esa\coth esa
    -1}{\sinh^2 esa} -\frac{1}{3} esa \coth esa \right] \, .\label{41}
\end{equation}
The complete formula for the effective action charge might be written
(substitution: $z:=esa$,
$h:=\case{m^2}{2ea}=\case{B^{\text{cr}}}{2B}$)

\begin{equation}
Q(h)=-\frac{1}{2a^2} \frac{\alpha}{\pi} \int\limits_0^{\text{i}\infty}
\frac{dz}{z} \text{e}^{-2hz} \left[ \frac{z\coth z-1}{\sinh^2 z}
  -\frac{1}{3} z\coth z \right]\, .\label{42}
\end{equation}
With some effort, the evaluation of the integral can be performed
analytically by standard means of dimensional regularization. Details
are given in App. A. The result is

%
%
%
%
\begin{eqnarray}
Q(h)=\frac{1}{2B^2}\frac{\alpha}{\pi}\biggl[&& \!\Bigl(\!
  2h^2\! -\case{1}{3}\Bigr) \psi (1\!+\!h) -h -3h^2-4h\ln \Gamma (h)
  \nonumber\\
&&+2h\ln 2\pi +\frac{1}{3} +4 \zeta '(-1,h) +\frac{1}{6h} \biggr] \,
  ,\label{43}
\end{eqnarray}
where $\psi$ denotes the logarithmic derivative of the
$\Gamma$-function and $\zeta '$ is the first derivative of the Hurwitz
Zeta function with respect to the first argument \cite{GR}.

For strong fields, the last term of eq.(\ref{43}) $\propto\case{1}{6h}
\propto |\bbox{B}|$ dominates the expression in the square
brackets. Hence, the effective action charge decreases with 

\begin{equation}
Q(B)\simeq \frac{1}{6}\frac{\alpha}{\pi} \frac{1}{B_{\text{cr}}}
\frac{1}{B}\, ,\, \text{for}\quad B\to \infty \label{43nach}
\end{equation}
which supports the charge picture (fig. 1). 

\begin{figure}
\begin{center}
\epsfig{figure=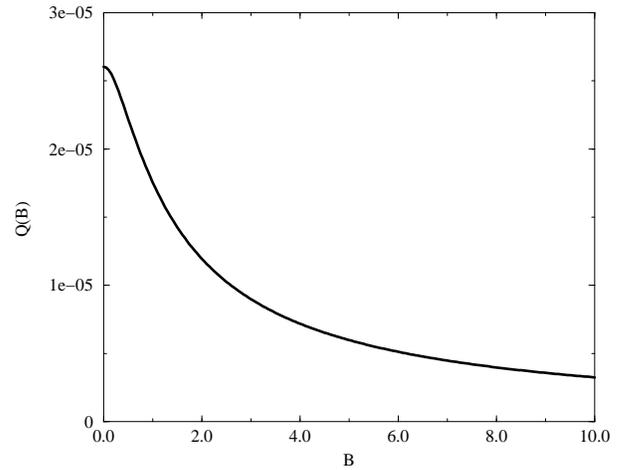,width=8cm}
\caption{Effective action charge $Q(B)=$ in units of
  $\frac{1}{m^4}$ versus magnetic field $B$ in units of the
  critical field strength $B_{\text{cr}}=\frac{m^2}{e}$.}
\end{center}
\end{figure}

The contraction of the energy-momentum tensor VEV may be cast into the
form

\begin{equation}
\langle T^{\mu\nu}\rangle \bar{k}_\mu\bar{k}_\nu =\bbox{B}^2-
(\bbox{B\cdot \hat{k}})^2+{\cal O}(\alpha) =B^2 \sin^2 \theta +{\cal
  O}(\alpha)\, ,\label{45}
\end{equation}
where $\theta$ measures the angle between the $\bbox{B}$-field and the
propagation direction.

Finally, the light cone condition eq.(\ref{25}) yields for arbitrary
background fields consistent with the one-loop approximation ($h=
\frac{B_{\text{cr}}}{2B}$) 

%
%
%
%
\begin{eqnarray}
v^2=1- &&\frac{\alpha}{\pi} \frac{\sin^2 \theta}{2} \nonumber\\
&&\times\biggl[\!\Bigl(
  \case{B_{\text{cr}}^2}{2B^2}\!-\!\case{1}{3}\Bigr) \psi (1\!+
  \!\case{B_{\text{cr}}}{2B})-\case{2B_{\text{cr}}}{B}\ln \Gamma
  (\case{B_{\text{cr}}}{2B}) -\case{3B_{\text{cr}}^2}{4B^2}  
  \nonumber\\
&&\quad\!-\case{B_{\text{cr}}}{2B}+\case{B_{\text{cr}}}{B}\ln 2\pi
  +\!\frac{1}{3}\!+4 \zeta '(-1,\case{B_{\text{cr}}}{2B})
  +\case{B}{3B_{\text{cr}}}  \biggr] .\nonumber\\
&&\label{46}
\end{eqnarray}
The first derivative of the Hurwitz Zeta function at $-1$ can be
related to the generalized $\Gamma$-function of first order $\Gamma_1$
\cite{dittgiesU}

\begin{equation}
\zeta '(-1,h)=-h\ln h +\ln \Gamma_1 (1+h) - L_1 +\frac{1}{12}\,
,\label{44}
\end{equation}
where $L_1=0.248\,754\,477\dots$ is a pure number and can be obtained
from the Raabe integral \cite{bendersky}.

%
%
Using eq.(\ref{44}), one can show that eq.(\ref{46}) is identical to
the findings of Tsai and Erber \cite{tsai1}. Equation (\ref{46}) is
plotted in fig. 2. Although the velocity shift increases proportional
to the magnetic field for large $B$, the total amount of the velocity
shift remains comparably small 

\begin{figure}
\begin{center}
\epsfig{figure=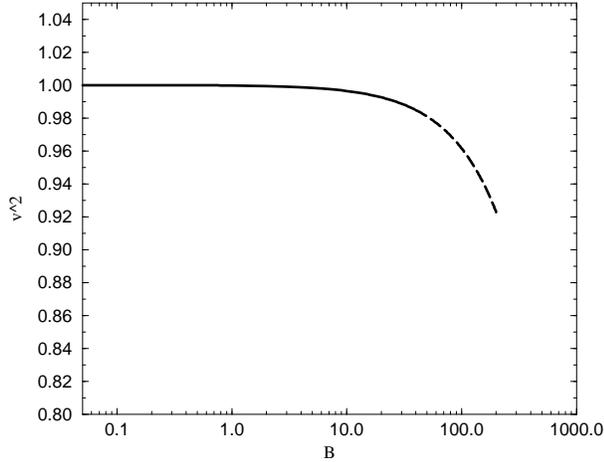,width=8cm}
\caption{Square velocity $v^2$ versus magnetic field $B$ in units of the
  critical field strength $B_{\text{cr}}=\frac{m^2}{e}$. The dashed
  curve indicates the region where two-loop corrections become important.}
\end{center}
\end{figure}

\begin{equation}
\delta v\simeq 9.58..\cdot 10^{-5}\quad \text{at}\quad
B=B_{\text{cr}}=\frac{m^2}{e} \label{46nach}
\end{equation}
for strong $B$-fields consistent with the one-loop approximation,
i.e., $\frac{B}{B_{\text{cr}}}<\frac{\pi}{\alpha}\simeq 430$. Taking
higher-order loop calculations into account, we expect a stronger
decrease of $Q(B)$ for large $B$ in order to let $Q \langle T_{\mu\nu}
\rangle \bar{k}_\mu \bar{k}_\nu$ be bounded.

\subsection{Casimir Vacua (Scharnhorst effect)}
One curious result regarding vacuum induced velocity shifts is
the possibility of superluminal phase and group velocities. As
mentioned above, e.g., Casimir vacua can create positive velocity
shifts, since a negative shift of the zero point energy is permitted.
For the configuration of perfectly conducting parallel plates of
distance $a$, $\langle T^{\mu\nu}\rangle$ is found to be \cite{brown1}

\begin{equation}
\langle T^{\mu\nu}\rangle =\frac{\pi^2}{720a^4} 
\left( \begin{array}{cccc}
       -1&  &  &  \\
         & 1&  &  \\
         &  & 1&  \\
         &  &  &-3 \end{array} \right) \, ,\label{47}
\end{equation}
%
%
where the symmetry axis points along the 3-direction.  The effective
action charge has to be evaluated in the zero-field limit. In
concordance with experimental facilities, the plate separation $a$ is
treated as a macroscopic parameter ($a\propto \mu m$); otherwise, we
would violate the soft-photon approximation, since the photon
wavelength $\lambda$ has to obey $\lambda\ll a$ to validate the
concept of treating the Casimir region as a (macroscopic) medium.

The magnitude of $a$ implies that we can neglect the $a$-dependence of
$Q$ which is exponentially suppressed by $ma\gg 1$ (this point will
become clearer in the following section). Here, Q is simply given by
eq.(\ref{36vor})

\begin{equation}
Q=c_1+c_2=\frac{2\alpha^2}{45m^4} \left(11+\frac{1955}{36}
  \frac{\alpha}{\pi} \right) \label{48}
\end{equation}
from which directly follows using eqs.(\ref{47}) and (\ref{25})

\begin{equation}
v=1+\frac{1}{(90)^2} \frac{\alpha^2}{m^4}\left(11+\frac{1955}{36}
  \frac{\alpha}{\pi} \right)\frac{\pi^2}{ a^4} \label{49}
\end{equation}
for a propagation perpendicular to the plates (a parallel propagation
will, of course, not be modified).

Equation (\ref{49}) represents the two-loop corrected version of
Scharnhorst's formula \cite{scharnhorst1,barton1}. Note that the
two-loop correction enhances the velocity shift. As was recently found
by Kong and Ravndal \cite{kong}, the radiative correction to the
Casimir energy is of order $\frac{\alpha^2}{m^4 a^8}$:

%
%
\begin{equation}
\langle T^{00}\rangle\equiv u=-\frac{\pi^2}{720 a^4}-
\frac{11}{(90)^2\cdot 30\cdot 16} \frac{\pi^4\alpha^2}{m^4 a^8} \,
.\label{50}
\end{equation}
At the two-loop level of eq.(\ref{49}), this correction can obviously
be neglected. Even three-loop contributions in $Q$ would be more
important. But it is interesting to note that this correction also
contributes positively to $v$.

\subsection{Finite Temperature}
In the remaining sections, we reveal the manifold features of
temperature induced velocity shifts. Unlike the Scharnhorst effect, we
do not recognize a principal obstacle against measurability here, and
the results allow for an immediate physical interpretation. The
following calculations are restricted to the one-loop level. 

We begin with the one-loop correction to the effective QED Lagrangian
at finite temperature which can be decomposed according to

\begin{equation}
{\cal L}_{\text{c}}(x,y,T)={\cal L}_{\text{c}}(x,y,T=0)+\Delta {\cal
  L} (x,y,T)\, ,\label{51}
\end{equation}
whereby ${\cal L}_{\text{c}}(x,y,T=0)$ denotes the usual
zero-tem\-pera\-ture Lagrangian eq.(\ref{37}). 

For purely magnetic fields, $\Delta {\cal L} (x,y,T)$ was calculated
by Dittrich \cite{dittrich1}

\begin{eqnarray}
%
%
\Delta {\cal L}(B,T)=-\frac{\sqrt{\pi}}{4\pi^2}
  \int\limits_0^{\text{i}\infty}&& \frac{ds}{s^{\case{5}{2}}}
  \text{e}^{-m^2 s}esB\cot esB \label{52}\\
&& \times T\left[ \Theta_2(0,4\pi\text{i}sT^2) -\frac{1}{2T\sqrt{\pi
  s}} \right]\, .\nonumber
\end{eqnarray}
The Jacobi $\Theta$-function is defined by \cite{GR}

\begin{equation}
\Theta_2(0,-q)=\sum_{n=-\infty}^\infty \exp\Bigl( -\text{i}q\bigl(
n+\case{1}{2} \bigr)^2 \Bigr)\, .\label{53}
\end{equation}
The effective action charge can be decomposed similarly to
eq.(\ref{51}) into

\begin{eqnarray}
%
%
Q(x,y,T)&=&Q(x,y,T=0)+\Delta Q(x,y,T)\, \label{54}\\
&=&\case{1}{2}\bbox{\nabla}^2{\cal L}_{\text{c}}(x,y,T=0)+\case{1}{2}
\bbox{\nabla}^2\Delta {\cal L} (x,y,T)\, .\nonumber
\end{eqnarray}
$Q(x,y,T=0)$ clearly corresponds to the zero-tem\-pera\-ture case as
treated above. 

%
%
Since we have to differentiate with respect to $x$ {\em and} $y$, it
is not sufficient for the calculation of $\Delta Q$ to consider
magnetic fields only in eq.(\ref{52}). Not until we have carried out
the Laplacian are we allowed to set $\bbox{E}=0$. Indeed, we have to
take this limit $\bbox{E}\to 0$ in the end, because the principle of
equilibrium thermodynamics would otherwise be violated. Besides, the
above-mentioned assumption of passivity of the vacuum is only
fulfilled for $\bbox{E}=0$. The appropriate expression is simply
obtained by replacing

\begin{mathletters}
\label{55}
\begin{equation}
esB\cot esB\label{55a}
\end{equation}
by the gauge and Lorentz invariant terms

\begin{equation}
(esa)(esb)\coth(esa)\cot(esb) \label{55b}
\end{equation}
\end{mathletters}

\noindent
in analogy to eq.(\ref{37}). Again, we make use of the
ad\-van\-tage\-ous coordinates $a,b$ in field space defined in
eqs.(\ref{38}). The result of the differentiation was already found in
eq.(\ref{40}); hence we obtain for the temperature induced effective
action charge for purely magnetic fields ($a=B,b=0$)

\begin{eqnarray}
\Delta&&\!Q(B,T)\nonumber\\
&&=\!-\frac{\sqrt{\pi}}{a^2}\frac{\alpha}{\pi}
  \!\!\int\limits_0^{\text{i}\infty}\!\! \frac{ds}{\sqrt{s}}
  \text{e}^{-m^2 s} { \left[ \frac{esa\coth esa
  -\! 1}{\sinh^2 esa} -\frac{esa}{3}  \coth esa \right]}\nonumber\\
&&\qquad\qquad\qquad \times T\left( \Theta_2(0,4\pi\text{i}sT^2)
  -\case{1}{\sqrt{\pi s} 2T}\right) \nonumber\\
&&=-\frac{\alpha}{\pi}\frac{1}{a^2}\!\int\limits_0^{\text{i}\infty}
\!\! \frac{ds}{s} \text{e}^{-m^2 s}{\left[ \frac{esa\coth
  esa-\! 1}{\sinh^2 esa} -\frac{esa}{3}  \coth esa \right]} \nonumber\\
&&\qquad\qquad\qquad\qquad\quad\times \sum_{n=1}^\infty
\text{e}^{-\text{i}\pi n} \text{e}^{-\frac{n^2}{4T^2 s}}\, .\label{56}
\end{eqnarray}
In the last line, we made use of the identity \cite{dittrich1}

\begin{equation}
\Theta_2(0,4\pi\text{i}sT^2)=\frac{1}{\sqrt{\pi s} 2T}\left( 1+ 2
  \sum_{n=1}^\infty \text{e}^{-\text{i}\pi n}
  \text{e}^{-\frac{n^2}{4T^2 s}}\right) .\label{57}
\end{equation}
%
%
Our task is to evaluate eq.(\ref{56}) in the various limits. First, we
consider pure temperature phenomena with vanishing field strength. The
temperature dependent part of the effective action charge reduces to

\begin{eqnarray}
\Delta Q(B=0,T)&&=2\frac{22}{45}\alpha^2 \int\limits_0^{\text{i}\infty}
  ds \,s \text{e}^{-m^2 s}\sum_{n=1}^\infty \text{e}^{-\text{i}\pi n}
  \text{e}^{-\frac{n^2}{4T^2 s}} \nonumber\\
=\frac{22}{45}\frac{\alpha^2}{m^4}&& \sum_{n=1}^\infty (-1)^n \left(
  \frac{m}{T} n\right)^2 K_2\left(\case{m}{T} n\right)\, ,\label{58}
\end{eqnarray}
whereby we have taken advantage of the representation

\begin{equation}
2\left(\case{\mu}{2}\right)^\nu K_\nu (\mu)=\int\limits_0^\infty du\,
u^{\nu-1} \exp \left( -u -\case{\mu^2}{4u} \right) \label{59}
\end{equation}
for the modified Bessel function and have rotated the contour. 

For low temperature, we may use the asymptotic expansion of $K_2(x)$
for $x\gg 1$

\begin{equation}
K_2 (x)=\sqrt{\frac{\pi}{2x}} \text{e}^{-x} \left(1+{\cal O} \left(
    \case{1}{x} \right) \right)\, .\label{60}
\end{equation}
In this limit, we find

\begin{eqnarray}
\Delta Q(B=0,T\to 0)&\simeq&-\frac{22}{45}\frac{\alpha^2}{m^4}
  \sqrt{\frac{\pi}{2}} \left(\frac{m}{T} \right)^{\frac{3}{2}}
  \text{e}^{-\frac{m}{T}} \nonumber\\
&\to& 0^-\, .\label{61}
\end{eqnarray}
Hence, the effective action charge is perfectly described by
$Q(B=0,T=0)=c_1+c_2$ eq.(\ref{36vor}) in this limit, while the
influence of temperature on $Q$ vanishes as it should. (Note that in
the case of Scharnhorst's effect a similar term $\Delta Q(B=0,ma\gg
1)$ also vanishes by drawing the analogy $T\propto \frac{1}{a}$.)

Next, we investigate the high-temperature li\-mit $T/m\gg 1$ of
eq.(\ref{58}). The calculation is, however, much more involved, so we
simply state the result

\begin{equation}
\Delta Q(T\gg m)=-\frac{22}{45} \frac{\alpha^2}{m^4} \left[
  1-\frac{k_1}{4} \frac{m^4}{T^4} +{\cal O}\left(\frac{m^6}{T^6}
  \right) \right]\, ,\label{61nach}
\end{equation}
where $k_1=0.123\,749\,077\,470\dots=$const.. The interested reader is
referred to App. B. 

Therefore, we arrive at the remarkable result that the complete
effective action charge 

\begin{eqnarray}
Q(T\gg m)&=&Q(T=0)+\Delta Q(T\gg m) \nonumber\\
&=&\frac{11}{90} k_1 \frac{\alpha^2}{T^4} +{\cal O}\left(
  \case{m^2}{T^6} \right) \label{62}
\end{eqnarray}
decreases rapidly $\propto 1/T^4$. $Q(B=0,T)$ is plotted in fig. 3.
The influence of temperature causes the effective action charge to
evaporate. Numerical results astonishingly indicate that eq.(\ref{62})
is already a reasonable approximation for $\frac{m}{T}\simeq 1.4$
(error$\leq 5\%$) where real $e^+ e^-$-pair creation is energetically
impossible and the vacuum is essentially modified by a photon gas.
This excludes the interpretation that eq.(\ref{62}) is a pure
threshold effect of pair production. 

\begin{figure}
\begin{center}
\epsfig{figure=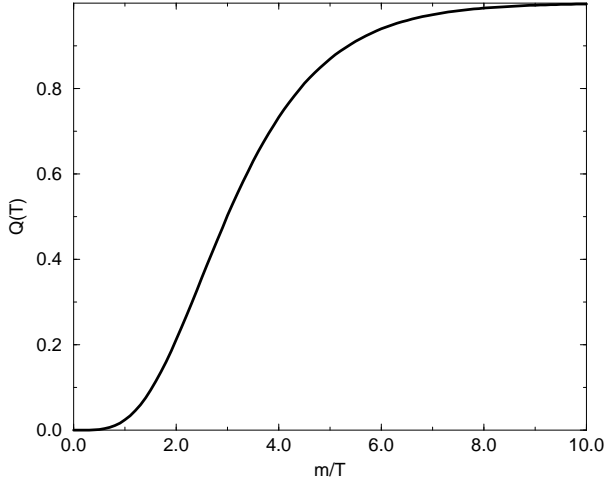,width=8cm}
\caption{Effective action charge $Q(T)$  $=Q(B=0,T=0)$ $+
  \Delta Q(T)$ in units of $\frac{22}{45}\frac{\alpha^2}{m^4}$; for
  high temperature, $Q(T)$ decreases proportional to $\frac{1}{T^4}$.}
\end{center}
\end{figure}

To complete the high temperature/zero field analysis of the light cone
condition we need the VEV of the energy-momentum tensor which is given
by (see, e.g., \cite{bailin})

\begin{equation}
\langle T_{\mu\nu} \rangle_T=\frac{\pi^2}{90} \left( N_{\text{B}}
  +\case{7}{8} N_{\text{F}} \right) T^4\, \text{diag} (3,1,1,1)\,
  .\label{63}
\end{equation}
The integer variables $N_{\text{B}}$ and $N_{\text{F}}$ denote the
number of bosonic and fermionic degrees of freedom at a given
temperature. For QED, we obtain

\begin{mathletters}
\label{64}
\begin{eqnarray}
N_{\text{B}}=2\, &,&\quad N_{\text{F}}=0\label{64a}\\
&&\text{for}\quad T\ll m\quad \text{(photon gas)} \nonumber\\
N_{\text{B}}=2\, &,&\quad N_{\text{F}}=4\label{64b}\\
&&\text{for}\quad T\gg m\quad \text{(photon + ultrarelativistic}
\nonumber\\ 
&&\qquad\qquad\qquad\quad e^+\,\text{and}\,\, e^-\,\text{fermion gas)}
\nonumber 
\end{eqnarray}
\end{mathletters}
It is appropriate to employ eq.(\ref{27}) for the light cone
condition. Using our findings in
%
%
%
%
eqs.[(\ref{61}), (\ref{36vor}), (\ref{63}), (\ref{64a})], we recover
the well-known result \cite{latorre} for low temperature

\begin{equation}
v=1-\frac{44\pi^2}{2025} \alpha^2 \frac{T^4}{m^4} \label{65}
\end{equation}
%
%
which according to fig. 3 is valid for $T/m< 0.16$ (error$\leq 5\%$)
($T<10^9$K). Substituting eq.(\ref{62}) into eq.(\ref{27}) and using
eqs.[(\ref{63}), (\ref{64b})] for $T\gg m$, we finally arrive at the
velocity of soft photons moving in a photon and ultrarelativistic $e^+
e^-$ gas

\begin{eqnarray}
v&=&1-\frac{121}{8100} k_1 \pi^2 \alpha^2+{\cal O}\!\left(\!
  \case{m^2}{T^2}\!\right) \nonumber\\
&=&1-9.72..\cdot 10^{-7}+{\cal O}\!\left(\!\case{m^2}{T^2}\!\right)
  =\text{const.} +{\cal O}\!\left(\!\case{m^2}{T^2}\!\right) \,
  .\label{66} 
\end{eqnarray}
In eqs.[(\ref{65}), (\ref{66})], we found that the velocity shift increases
proportional to $T^4$ for low temperature but approaches a constant
value in the high-temperature limit. This can be understood in terms
of the effective action charge which evaporates sufficiently fast
compared to the increase of the energy-momentum tensor VEV. Obviously,
the shift described by eq.(\ref{66}) remains small; therefore the
deviation from the vacuum velocity does not become seriously important
(e.g., for the construction of cosmological models). However, one
should keep in mind that, if the temperature exceeds the masses of
further charged particles, each particle will contribute additively to
$Q$ and will increase the respective number of degrees of freedom
$N_{\text{B}}$ or $N_{\text{F}}$.

\subsection{Casimir Vacua at Finite Temperature}
The combination of thermal and Casimir phenomena is in itself
worthwhile studying, because both effects enter the formalism via
boundary conditions but lead to opposite results. In the following, we
want to investigate where and why the respective effect dominates the
velocity shift. The determining order parameter is the dimensionless
combination $Ta$. Nevertheless, the plate separation $a$ has to be
considered as a macroscopic quantity ($a\simeq \mu m$).

First, we consider the low-temperature region. According to Brown and
Maclay \cite{brown1}, the VEV of $T^{\mu\nu}$ depending on $a$ and $T$
is given by

\begin{eqnarray}
\langle T^{00}\rangle_T^a &=& -\frac{\pi^2}{720}\frac{1}{a^4}
  +\frac{\zeta (3)}{\pi^2} \frac{T^3}{a}\, , \nonumber\\
\langle T^{33}\rangle_T^a &=& -\frac{\pi^2}{240}\frac{1}{a^4}\,
  ,\quad\text{for}\quad Ta\to 0\, ,\label{67}
\end{eqnarray}
for the parallel plate configuration ($\zeta (3)=1.202056\dots$). The
light cone condition eq.(\ref{25}) for a propagation perpendicular to
the plates ($\bar{k}^\mu=(v,0,0,1)$) yields

\begin{equation}
v=1+\frac{1}{(90)^2} \frac{\alpha^2}{m^4}\left(\! 11+\case{1955}{36}
\case{\alpha}{\pi}\right)\! \frac{\pi^2}{a^4} \left(\!
1-\case{180\zeta (3)}{\pi^4} (Ta)^3\right) \, .\label{68}
\end{equation}
In the low-$T$ limit, the $(Ta)^3$-term can be neglected and we only
rediscover Scharnhorst's result. But we do not find an additional
velocity shift proportional to $T^4$ which could have been expected
from eq.(\ref{65}). This clearly arises from the fact that none of
the (quantized) perpendicular modes can be excited at
low temperature. The $(Ta)^3$-term in eq.(\ref{68}) will become
important for $Ta={\cal O}(1)$, i.e., $T>2000$K for $a\simeq
\mu$m. This shows that the Scharnhorst effect is not sensitive to
temperature perturbations.

For increasing temperature, we encounter an intermediate temperature
region characterized by the condition $1\ll Ta\ll ma$ which
corresponds to $0.2$eV$\ll T\ll 0.5 $MeV. This implies that $Q=Q(T=0)$
is a justified approximation and the thermal contribution of an $e^+
e^-$ gas does not have to be taken into account. 

Using further results of Brown and Maclay \cite{brown1}

\begin{eqnarray}
\langle T^{00}\rangle_T^a &=& \frac{\pi^2}{15} T^4\, \label{69}\\  
\langle T^{33}\rangle_T^a &=& \frac{\pi^2}{45}T^4 +\frac{\zeta
  (3)}{4 \pi} \frac{T}{a^3}\, ,\quad\text{for}\quad Ta\gg 1\, ,\nonumber
\end{eqnarray}
we find 

\begin{equation}
v=1-\frac{4\pi^2}{(45)^2} \frac{\alpha^2}{m^4}\left(\!
  11+\case{1955}{36}\case{\alpha}{\pi}\right)\! T^4\left(\!
  1-\case{45\zeta (3)}{16\pi^3} \frac{1}{(Ta)^3}\right) \, .\label{70}
\end{equation}
In this limit, only the modifications caused by the black body
radiation become important. A term proportional to $1/a^4$ as a
consequence of certain missing zero-point fluctuations does not occur,
since higher (perpendicular) modes have been thermally excited.

For $T\gg m$, we will certainly recover eq.(\ref{66}) with negligible
$1/(Ta)^3$ Casimir corrections. Anyway, the concept of solid plates is
(at least experimentally) meaningless in this domain.

\subsection{Finite Temperature and Magnetic Fields}
For low temperature as well as for weak fields, thermal phenomena
decouple from magnetic vacuum modifications, because the effective
action charge is not sensitive to weak influences. The velocity shifts
can simply be described by an addition of the respective
above-calculated ones. The only non-trivial interplay can be found in
the domain of strong fields in hot surroundings (e.g., hot neutron
stars). Our intension is to evaluate the thermal effective action
charge contribution given in eq.(\ref{56}) in this limit. Therefore,
we substitute $z=esa$ ($h=\frac{m^2}{2ea}=\frac{B_{\text{cr}}}{2B}$)

\begin{eqnarray}
\Delta Q(h,T)=-\frac{\alpha}{\pi}\frac{1}{a^2}\!
  \int\limits_0^{\text{i}\infty}\!\frac{dz}{z}\text{e}^{-2hz}&& \left[
  \frac{z\coth z-\!1}{\sinh^2 z} -\frac{1}{3} z\coth z \right]
  \nonumber\\
&& \sum_{n=1}^\infty (-1)^n \text{e}^{-\frac{ea}{4T^2}
  \frac{n^2}{z}}\, .\label{71}
\end{eqnarray}
In this representation, it is obvious that the integral is dominated
by small values of $z$ for weak fields ($h\gg 1$) and vice versa,
i.e., large $z$ for strong fields. We are interested in the latter, so
we expand the term in the square brackets for $z\gg 1$: $[\cdots]\to
-\case{1}{3} z$. Following the manipulations of
eqs.[(\ref{58}), (\ref{59})], we arrive at 

%
%
\begin{equation}
\Delta Q(T,B\gg B_{\text{cr}})=\frac{1}{3}\frac{\alpha}{\pi} \frac{1}{B^2}
\frac{B}{B_{\text{cr}}} \sum_{n=1}^\infty (-1)^n \frac{m}{T}n\, K_1 \left(
  \case{m}{T} n\right)\, .\label{72}
\end{equation}
%
%
Note that it was not necessary to impose any conditions on the
magnitude of $T$ to arrive at eq.(\ref{72}). But, as mentioned above,
the field-temperature phenomena decouple for $T\ll m$ due to the
asymptotic behaviour of $K_1(\case{m}{T} n)\propto\exp (-\case{m}{T})$
in eq.(\ref{72}); hence, $\Delta Q(\case{m}{T}\gg 1, B\gg B_{\text{cr}})\to
0$. Using similar techniques as applied in App. B, the
high-temperature limit of eq.(\ref{72}) can be determined. The result
for $T\gg m$ and $B\gg B_{\text{cr}}$ is

%
%
\begin{eqnarray}
\Delta Q(T,B)&=&\frac{1}{3}\frac{\alpha}{\pi}
  \frac{1}{B^2} \frac{B}{B_{\text{cr}}} \left(
  -\frac{1}{2}+\frac{1}{2}k_2\frac{m^2}{T^2} +{\cal 
  O}\left(\! \frac{m^4}{T^4}\!\right)\right) \nonumber\\
&=&-\frac{1}{6}\frac{\alpha}{\pi} \frac{1}{B^2}
  \frac{B}{B_{\text{cr}}} +\frac{1}{6}\frac{\alpha}{\pi} k_2
  \frac{e}{BT^2} +{\cal O}\left(\! \frac{m^2}{BT^4}\!\right)\,
  ,\nonumber\\
&&\label{73}
\end{eqnarray}
where $k_2=0.213\,139\,199\,408\dots=$const. (see eq.(\ref{B14}). To
obtain the complete effective action charge $Q$, we add the
strong-field contribution $Q(T=0)$ which was found in
eq.(\ref{43nach})

\begin{eqnarray}
Q(T\gg m,B\gg B_{\text{cr}})&=&\frac{1}{6}\frac{\alpha}{\pi} k_2
  \frac{e}{BT^2} +{\cal O}\left(\! \frac{m^2}{BT^4}\!\right)
  \nonumber\\
&=&\frac{2}{3}k_2 \frac{\alpha^2}{m^4} \frac{1}{\tilde{B}\tilde{T}^2}
  +{\cal O}\left(\! \frac{m^2}{BT^4}\!\right) \, ,\label{84}
\end{eqnarray}
where we have introduced the convenient dimensionless variables
$\tilde{B}=\frac{B}{B_{\text{cr}}}=\frac{eB}{m^2}$ and
$\tilde{T}=\frac{T}{m}$ which satisfy $\tilde{B},\tilde{T}\gg 1$.
Equation (\ref{84}) describes the same features of the effective
action charge which we have encountered in previous examples: $Q$ is
centred upon the origin in field space, decreases proportional to
$1/\tilde{B}$ and evaporates with increasing temperature. 

To calculate the velocity shift, we need the energy density which
consists of three parts

\begin{equation}
\langle T^{00}\rangle =\underbrace{\langle
  T^{00}\rangle_T^{B=0}}_{\text{eq}.(\ref{66})} +\underbrace{\langle
  T^{00}\rangle_{T=0}^{B}}_{\case{1}{2} B^2} +\Delta\langle
  T^{00}\rangle_T^{B} \label{85} 
\end{equation}
The last term of eq.(\ref{85}) is connected with the Lagrangian via
the free energy (density) according to 

\begin{eqnarray}
\Delta\langle T^{00}\rangle_T^{B}&=&F+TS=F-T\frac{\partial F}{\partial
  T} \nonumber\\
&=&-{\cal L}_T^B +T\frac{\partial {\cal L}^B_T}{\partial T} \,
  .\label{86}
\end{eqnarray}
The leading mixed contribution ${\cal L}_T^B$ to ${\cal L}$ is found
in ref.\cite{dittrich1}

\begin{equation}
{\cal L}_T^B=\frac{eB}{12}T^2 \quad\leadsto\quad \Delta\langle
T^{00}\rangle_T^{B} =\frac{eB}{12}T^2 \, .\label{87}
\end{equation}
We finally arrive at the polarization and propagation direction
averaged velocity shift for strong fields and high temperature 

\begin{eqnarray}
v&=&1-\frac{11\pi^2}{135}k_2 \alpha^2 \frac{\tilde{T}^2}{\tilde{B}}
  -\frac{k_2}{18\pi} \alpha \frac{\tilde{B}}{\tilde{T}^2}
  -\frac{k_2}{27}\alpha^2\, .\label{88}\\
&=&1-9.13..\cdot 10^{-6} \frac{\tilde{T}^2}{\tilde{B}}- 2.75..\cdot
  10^{-5} \frac{\tilde{B}}{\tilde{T}^2} -4.21..\cdot 10^{-7}\!
  .\nonumber
\end{eqnarray}
At $\tilde{T}^2/\tilde{B}=1.74\dots$, we find a minimal velocity
shift

\begin{equation}
|\delta v|\simeq 3.20\cdot 10^{-5}. \label{89}
\end{equation}
At the same time, this number approximately sets the scale of a
typical velocity shift for strong fields consistent with the one-loop
approximation. This is also confirmed by the result of
eq.(\ref{46nach}). 

\section{conclusions}
In this work, we studied light propagation in non-trivial QED vacua in
the geometric optics approximation. For any given QED effective
action describing soft photons, we derived the light cone condition
averaged over polarization states. This result generalizes the
``unified formula'' found by Latorre, Pascual and Tarrach
\cite{latorre} which turned out to be the low-energy limit of
our light cone condition. 

We furthermore clarified the connection between light velocity shifts
and the scale anomaly suggested by Shore \cite{shore}. Unfortunately,
our findings do not indicate an immediate connection hinting at deeper
physical grounds. 

Instead, the structure of the light cone condition suggests
introducing the intuitive physical picture of an effective action
charge $Q$ showing a localized profile in field space centred upon the
origin. This charge directly characterizes the properties of the
modified vacuum which are responsible for velocity shifts.

Within this conceptual framework, we analysed several modified QED
vacua and calculated the respective modified velocities. The inverse
velocities are equal to the refractive indices of the modified vacua
in the low-frequency limit. Hence, these velocities are phase as well
as group velocities -- the latter due to their independence of
frequency.  In the low-energy limit, we recovered all known results
which were already perfectly described by the ``unified formula''.

For arbitrary magnetic fields, we reproduced the findings of Tsai and
Erber \cite{tsai1} using our comparably simple formalism.

In the sequel, we calculated the next-to-leading order corrections to
the Scharnhorst effect. 

Finally, we investigated the influence of temperature on the velocity
shifts. The evaporation of the effective action charge turns out to be
the dominating effect in the high-temperature domain. It causes the
velocity shift to approach a constant value. Only when a strong
magnetic field is involved, does the light cone condition fail to
provide for a bound of the velocity shift in our examples. But we
expect higher-order loop corrections to stop an unbounded growth of
the velocity shift by inducing a faster decrease of the effective
action charge far from the origin in field space.

Referring to the light cone condition, the sign of the velocity shift
is in general determined by the sign of the effective action charge
{\em and} the vacuum energy density. However, up to now, we have not
been able to construct an example which exhibits a negative effective
action charge in QED. This might be a general characteristic of the
abelian theory. Indeed, the one-loop effective action of a covariant
constant chromomagnetic background field \cite{savvidy} (naively)
possesses a negative effective action charge.

We would like to conclude with the remarkable observation that parity
violating terms in the effective action proportional to $y^{2n+1},
n=0,1,2\dots$ will not contribute to the effective action charge in
the zero field limit, since the equation for $Q$ is of Poisson type.
Thus, e.g., the existence of dyons \cite{kovalevich} will not cause a
velocity shift in the weak field limit.

\appendix

\section{}
Our aim is to evaluate the integral of eq.(\ref{42})

\begin{equation}
I(h)=\int\limits_0^{\text{i}\infty} \frac{dz}{z} \text{e}^{-2hz}
  \left[ \frac{z\coth z-1}{\sinh^2 z} -\frac{1}{3} z\coth z \right]\,
  .\label{A1} 
\end{equation}
For this, we have to decompose it into simple parts which one can
handle by standard methods of dimensional regularization. Note that
the integral is convergent, since the prescription $h\to h-\text{i}
\epsilon$ is implicitly understood.

We begin with an integration by parts of the first term in square
brackets with respect to the sinh$^2$ in the denominator. This leads
to

\begin{eqnarray}
I(h)&=&\int\limits_0^{\text{i}\infty}dz\text{e}^{-2hz} \left[
  \frac{h}{z}\coth z-h\coth^2 z+\frac{1}{2z^2} \coth
  z\right. \nonumber\\
&&\qquad\qquad\quad\left. -\frac{1}{2z} \frac{1}{\sinh^2 z}
  -\frac{1}{3} \coth z \right] \, -\frac{1}{6} \label{A2}
\end{eqnarray}
The last three terms of the expression in square brackets are already
in a convenient shape. In the following, we thus consider only the
remaining first two terms. The strategy is similar: we extract a term
proportional to $1/\sinh^2 z$ and integrate by parts.

\begin{eqnarray}
I_1(h)&:=&h\int\limits_0^{\text{i}\infty}dz\text{e}^{-2hz} \left[
  \coth a\left( \frac{1}{z} -\coth z\right)\right] \nonumber\\
&=&h\int\limits_0^{\text{i}\infty}dz\text{e}^{-2hz} \left[ \frac{\cosh
  z\sinh z}{z} -\cosh^2 z \right] \frac{1}{\sinh^2 z} \nonumber\\
&=&h\int\limits_0^{\text{i}\infty}dz\text{e}^{-2hz}\!\left[\!\left(\!
  2h \!+\!\case{1}{z}\!\right) \coth z+ \left(\! h\!+\!\case{1}{z}
  \!\right) \sinh 2z \right.\nonumber\\
&&\qquad\qquad \left. -\left(\! \case{h}{z}\!+\!\case{1}{2z^2}\!+\!1
  \!\right) \cosh 2z -\left(\! \case{h}{z}\!+\!\case{1}{2z^2}\!+\!1
  \right) \right] \nonumber\\
&&\label{A3}
\end{eqnarray}
Inserting $I_1$ into eq.(\ref{A2}), we obtain the wanted types of
integrals. Each of these can be integrated separately by introducing
an extra factor of $z^\epsilon$ and rotating the contour onto the
positive real axis. At the end, the $1/\epsilon$-poles cancel and we
arrive at the result given in eq.(\ref{44}) in the limit $\epsilon\to
0$.

\section{}
In this appendix, we want to expand the infinite sum in eq.(\ref{58})
for small values of $\lambda:=\frac{m}{T}$ which corresponds to a high
temperature limit

\begin{equation}
S(\lambda):=\sum_{n=1}^\infty (-1)^n \left( \lambda n\right)^2
K_2(\lambda n)\, . \label{B1}
\end{equation}
Since the appearance of Bessel functions reflects the $R^3\times S^1$
topology which is the finite-temperature field theory space, the
techniques described in the following are certainly useful for further
finite-temperature applications.

The first step is to choose a representation of the modified Bessel
function that shows a simple dependence on the summation index
\cite{GR}

\begin{equation}
K_2(\lambda n)=\int\limits_0^\infty \text{e}^{-\lambda n\cosh t}\cosh
2t\, dt \, .\label{B2}
\end{equation}
Inserting eq.(\ref{B2}) into eq.(\ref{B1}), leads us to

\begin{equation}
S(\lambda)=\lambda^2 \int\limits_0^\infty dt\, \cosh
2t\sum_{n=1}^\infty n^2\, \text{e}^{-(\text{i}\pi +\lambda\cosh t)n}\,
.\label{B3} 
\end{equation}
By differentiating the geometric series $\sum_{n=0}^\infty
q^n=\frac{1}{1-q}$ twice with respect to $q$, we find the result for
the sum in eq.(\ref{B3})

\begin{equation}
\sum_{n=0}^\infty n^2 q^n=\frac{q(1+q)}{(1-q)^3}\, .\label{B4}
\end{equation}
Inserting eq.(\ref{B4}) into eq.(\ref{B3}) and decomposing the $\cosh
2t$ into $2\cosh^2 t -1$, we get

\begin{eqnarray}
S(\lambda)&=&\lambda^2\int\limits_\lambda^\infty
  \frac{dp}{\sqrt{p^2-\lambda^2}} \text{e}^{-p} 
  \frac{\left( 1-\text{e}^{-p}\right)}{\left( 1+\text{e}^{-p}
  \right)^3} \nonumber\\
&&-2\int\limits_\lambda^\infty \frac{p^2\,dp}{\sqrt{p^2-\lambda^2}}
  \text{e}^{-p} 
  \frac{\left( 1-\text{e}^{-p}\right)}{\left( 1+\text{e}^{-p}
  \right)^3}\,  \nonumber\\
&=&-2\int\limits_\lambda^\infty dp\, \sqrt{p^2-\lambda^2} \text{e}^{-p}
  \frac{\left( 1-\text{e}^{-p}\right)}{\left( 1+\text{e}^{-p}
  \right)^3}\,  \nonumber\\
&&-\lambda^2\int\limits_\lambda^\infty \frac{dp}{\sqrt{p^2-\lambda^2}}
  \text{e}^{-p}   \frac{\left( 1-\text{e}^{-p}\right)}{\left(
  1+\text{e}^{-p} \right)^3} \nonumber\\
&=:&J_1(\lambda)+\lambda^2\,J_2(\lambda)\, ,\label{B5}
\end{eqnarray}
where we have substituted $p:=\lambda\cosh t$. With some care, the
parameter integrals $J_1$ and $J_2$ can now be expanded. We have to
pay special attention to the process of taking the limit $\lambda\to
0$ for the $J$'s and their derivatives. We can circumvent possible
convergence problems at the lower bound by a repeated integration by
parts of the square root terms. Using the short form

\begin{equation}
(\%):= \frac{1}{p} \text{e}^{-p} \frac{\left(
  1-\text{e}^{-p}\right)}{\left( 1+\text{e}^{-p} \right)^3}\, ,
  \label{B6}
\end{equation}
the non-vanishing coefficients of the expansion up to order ${\cal
  O}(\lambda^5)$ can be expressed as

\begin{eqnarray}
J_1(0)&=&2\int\limits_0^\infty dp\, p^2 \, (\%)=1\, ,\label{B7}\\
J_1''(0)&=&2\int\limits_0^\infty dp\, p\frac{d}{dp}
  (\%)=-2\int\limits_0^\infty dp\, (\%)\, ,\label{B8}\\ 
J_1''''(0)&=&6\int\limits_0^\infty dp\, p\frac{d}{dp}\left(\frac{1}{p}
  \frac{d}{dp} (\%)\right) \, ,\label{B9}\\
J_2(0)&=&\int\limits_0^\infty dp\, (\%)\,=-2\, J_1''(0)\,
  ,\label{B10}\\
J_2''(0)&=&-\int\limits_0^\infty dp\, p\frac{d}{dp}\left(\frac{1}{p}
  \frac{d}{dp} (\%)\right) =-\frac{1}{6} J_1''''(0)\, ,\label{B11}
\end{eqnarray}
Finally, the Taylor expansion of eq.(\ref{B1}) reads

\begin{eqnarray}
S(\lambda)&=&J_1(0)+\left(\case{1}{2}J_1''(0)+J_2(0)\right)\lambda^2
  \nonumber\\ 
&&+\left(\case{1}{24} J_1''''(0)+\case{1}{2}J_2''(0)\right)\lambda^4
  +{\cal O}(\lambda^6) \nonumber\\
&=&1-\frac{k_1}{4} \lambda^4 +{\cal O}(\lambda^6)\, , \label{B12}
\end{eqnarray}
where the constant $k_1$ is defined by

\begin{eqnarray}
k_1&:=&-J_2''(0)\equiv\frac{1}{6} J_1''''(0)=\int\limits_0^\infty dp\,
  p\frac{d}{dp}\left(\frac{1}{p} \frac{d}{dp} (\%)\right) \nonumber\\
&=&0.123\,749\,077\,479\dots \, .\label{B13}
\end{eqnarray}
The constant $k_2$ that appears in the calculation of the effective
action charge for high temperature and strong fields is obtained by
similar techniques. Its integral representation is (accidentally)
equal to $J_2(0)$

\begin{eqnarray}
k_2&:=&J_2(0)=\int\limits_0^\infty\frac{dp}{p} \text{e}^{-p}
  \frac{\left( 1-\text{e}^{-p}\right)}{\left( 1+\text{e}^{-p}
  \right)^3} \nonumber\\
&=&0.213\,139\,199\,408\dots\, . \label{B14}
\end{eqnarray}

\end{document}